\documentstyle[12pt,epsf]{article}
\begin{document}
\baselineskip=20pt plus 1pt minus 1pt
\tolerance=1500
\thispagestyle{empty}
\vskip1cm
\vglue1cm
\begin{center}
{\large{\bf TEMPERATURE DEPENDENCE OF HEAVY MESON MASSES}}
\end{center}
\vskip7mm
\begin{center}
F.S. NAVARRA, C.A.A. NUNES \\
\vskip3mm
Nuclear Theory and Elementary Particle Phenomenology Group\\
Instituto de F\'\i sica da Universidade de S\~ao Paulo,\\
Caixa Postal 66318, 05389-970 S\~ao Paulo, Brazil\\
\vskip5mm
\end{center}
\vskip15mm
\baselineskip=26pt plus1pt minus1pt

\vspace{2cm}

\noindent {\bf Abstract:}Using a previously derived QCD effective
hamiltonian
we find the masses of heavy quarkonia states. Non perturbative effects are
included through temperature dependent gluonic condensates. We find that
even
a moderate change in these condensates in a hot hadronic environment (below
the deconfining transition) is sufficient to significantly change the heavy
meson masses.
\vfill
\eject

\noindent
\vskip7mm

The study of hadronic matter at high temperatures and densities has
direct relevance for heavy-ion experiments. Apart from the search of
quark gluon plasma, serious attention has been given to signatures of a
hot system composed by hadrons below the deconfining transition. Among
these signatures, changes in the masses due to medium effects play a
major role. They have been extensively studied with the
Nambu-Jona-Lasinio model~[1], with non-relativistic potential
models~[2] , with QCD sum-rules~[3] and in lattice QCD~[4].

The experimental detection of medium effects on the masses is a very
difficult problem. On the other hand, from the theoretical point of
view the situation is not quite clear. Some calculations predict a
significant decrease ($\sim 400$~MeV) of the $\eta_c$, $J/\psi$ and
$\psi'$ masses. Some others suggest that the masses stay constant. We
will argue that they may increase.

The approach to this problem adopted in the present work
 is in many aspects similar to QCD sum-rules at finite
temperature. In particular, our results for the heavy quarkonium
spectrum depend primarily on perturbation theory (which is related to
the value of $\alpha_s$) and on the gluon condensate. The same
conclusion is found in ref.~[3]. The main difference between this work
and the above mentioned spectrum calculations is that we give special
emphasys to vacuum changes with temperature.

It has been known since the late seventies that the QCD (physical)
vacuum is full of soft gluons or, equivalently, contains chromoelectric
(E)
and magnetic (B) fields. Such a state, $|\Omega\rangle$, has lower energy
than a state without any fields, the perturbative vacuum, $|0\rangle$.
This picture is supported by the existence of non-vanishing gluon
condensates, i.e.,
\begin{equation}
\langle\Omega | \, \frac{\alpha_s}{\pi} \; F_{\mu\nu} \, F^{\mu\nu}
   |\Omega\rangle = \phi^2 \neq 0
\label{eq.1}
\end{equation}

where $F_{\mu\nu}$ is the usual QCD field tensor.

 At increasing
temperatures, lattice calculations predict a phase transition to a
deconfined phase. In terms of the vacuum state this transition can be
interpreted as the passage from the non-perturbative (or physical) to
the perturbative vacuum, i.e., $| \Omega \rangle \rightarrow
|0\rangle$. Roughly speaking, one can say that the background soft
gluon fields would then ``boil away'', implying that $\phi^2
\rightarrow 0$.  In fact some specific lattice calculations~{[5]}
suggest that even above the deconfining phase transition there may be
non-vanishing gluon condensates, but they are not yet conclusive and
therefore from lattice simulations we cannot extract the temperature
dependence of the gluon condensate over a wide range of temperatures.
Model calculations using chiral perturbation theory~{[6]} or dilaton
fields~{[7]} come to the conclusion that $\phi^2$ stays almost
constant until the phase transition temperature $T_c$ and then starts
to drop faster going to zero very slowly.

We will now shortly describe the basic elements of our ``effective QCD''
applied to the study of heavy quarkonium.
Our meson states incorporate the background soft gluon fields .
Together with the usual quark anti-quark states, $|q\overline
q\,\rangle$, we will have also $|q \, E \, \overline q\,\rangle$ or $|q
\, B \, \overline q \, \rangle$. In the first case the quark and
anti-quark are in a color singlet representation and we call it a
singlet state. In the second case, because of the interaction with the
vacuum, the combination $q \, E \, \overline q$ (or $q \, B \,
\overline q \ldots$) is a color singlet but the quark anti-quark pair
is in a color octect representation. We call therefore these states
octect states.

In order to ensure the gauge invariance of our calculations we will
make our meson states gauge invariant by construction. This can be done
with the help of the color transport operator
\begin{equation}
T_{ab}\left( \vec x_2, \vec x_1 \right) = P \, \exp \left(
    - i g \int^{\vec x_2,t}_{\vec x_1, t} dx^\mu \, A_\mu
     \right)_{ab}  \quad .
\label{eq.2}
\end{equation}
The path ordered exponential (denoted by $P\,\exp$) of the background
gluon field $A_\mu(x)$ transports along a straight line the color index
$b$ (of the fundamental representation of the gauge group) at position
$\vec x_1$ to index $a$ at position $\vec x_2$. This operator was
introduced in this context by Schaden and Glazek~{[8]} and extensively
used later on by Nunes~{[9]}. It is easy to show that canonical
anti-commutation relations for the quark and anti-quark operators imply
that the singlet meson basis states defined below are orthonormal.

Considering all that was said above we can write now our heavy meson basis
states explicitly. For simplicity we restrict ourselves to the
pseudoscalar mesons, which are then
\begin{eqnarray}
|S\rangle & = & \frac{1}{\sqrt{6}} \sum_{ab;\alpha}
   u^{\dagger\alpha}_a \left( \vec x_2 \right) \,
    T_{ab}\left( \vec x_2, \vec x_1 \right) \,
     v^\alpha_a\left( \vec x_1 \right) |\Omega\rangle
\label{eq.3} \\[0.3cm]
|B\rangle & = & \sum_{a,b,c;\alpha,\beta} \; \frac{g}{\pi\phi} \;
   u^{\dagger\alpha}_a \left( \vec x_2 \right) \,
          \vec\sigma^{\alpha\beta} \cdot \vec B_{ab} \, T_{bc}\left(
     \vec x_2, \vec x_1 \right) \, v^\beta_c\left( \vec x_1 \right)
      |\Omega\rangle
\label{eq.4} \\[0.3cm]
|E_1\rangle & = & \sum_{a,b,c;\alpha} \; \frac{\sqrt{3g}}{\pi\phi} \;
   u^{\dagger\alpha}_a \left( \vec x_2 \right) \,
          \vec E_{ab} \cdot \left( \vec x_2 - \vec x_1 \right) \,
      T_{bc} \left( \vec x_2, \vec x_1 \right) \,
       v^\alpha_c\left( \vec x_1 \right)  |\Omega\rangle
\label{eq.5} \\[0.3cm]
|E_2\rangle & = & \sum_{a,b,c;\alpha,\beta} \;
  \frac{i\,\sqrt{3g}}{\sqrt{2\pi} \, \phi} \;
    u^{\dagger\alpha}_a \left( \vec x_2 \right) \,
           \vec E_{ab} \cdot \left( \vec\sigma^{\alpha\beta}
      \times  \left( \vec x_2 - \vec x_1  \right)\right) \,
        T_{bc} \left( \vec x_2, \vec x_1 \right) \,
         v^\beta_c\left( \vec x_1 \right)  |\Omega\rangle  \quad .
\label{eq.6}
\end{eqnarray}

The pseudoscalar meson $\eta$ can be well represented by a linear
combination of the above basis states
\begin{equation}
| \eta \rangle = \sum_{M=S,E_1,E_2,B} \,
   \int_{1,2} \psi_M (2,1)\,|2,1\rangle_M  \quad .
\label{eq.6a}
\end{equation}

The interaction between heavy quarks is described by the QCD Lagrangian
with two simplifying approximations: a)~non relativistic limit with the
inverse heavy quark mass expansion up to first order and \
b)~separation of the gluon fields into classical background
nonperturbative fields (which will later give rise to the condensates)
and quantum high momentum fields. Expansions involving both types of
fields will include only lower order terms because we will consider
only the lowest order gluon condensates and also because higher powers
of the quantum high momentum fields will couple only in the
perturbative regime and can be neglected since $\alpha_s$ is small for
high momentum couplings.

Denoting the quark fields by $\psi$ and gluon fields by $V_\mu$ the QCD
Lagrangian is written as
\begin{equation}
{\cal L}_{QCD} \; = \; - \, \frac{1}{4} \; F^a_{\mu\nu} \,
  F^{\mu\nu}_a + \overline\psi \left( i\not\!\partial + g\,
   T^a \!\not\!V_a \right) \psi - m\overline \psi\psi \quad .
\label{eq.7}
\end{equation}
We make then a Foldy-Wouthuysen transformation in the quark fields
\begin{equation}
 \begin{array}{l}
   \psi \longrightarrow \exp\left( i\vec\gamma \cdot \vec D/2m
     \right) \psi \\
   \overline\psi \longrightarrow \overline\psi \exp\left(
      - \,i\vec\gamma \cdot \vec D/2m \right)
 \end{array}
\label{eq.8}
\end{equation}
obtaining a non-relativistic Lagrangian
\begin{equation}
{\cal L}_{NRQCD} \; = \; - \, \frac{1}{4} \; F^a_{\mu\nu} \,
  F^{\mu\nu}_a + \overline\psi \left( i \gamma^0 D_0 - m \right)
   \psi + \frac{1}{2m} \; \overline\psi \, \vec D^2 \, \psi +
    \frac{1}{2m} \; \overline\psi \, g\, \vec\Sigma \cdot \vec B \,
     \psi
\label{eq.9}
\end{equation}
where \ $\vec D = \vec\partial - ig\vec V$ and $\vec\Sigma = \left(
\begin{array}{cc} \vec\sigma & 0 \\ 0 & \vec\sigma \end{array} \right)$
\ does not couple upper and lower spinor components. We next separate
the gluon field in background ($A_\mu$) and quantum ($Q_\mu$) fields
\begin{equation}
V^a_\mu = A^a_\mu + Q^a_\mu  \quad . \label{eq.10}
\end{equation}

We choose the Coulomb background gauge for the quantum fields
\begin{equation}
D_i \, Q^i = 0    \label{eq.11}
\end{equation}
where $D_\mu Q_\nu = \partial_\mu  Q_{\nu a} + g f^{abc} A_{\mu b}
Q_{\nu c}$. The background fields are defined in a modified Schwinger
gauge~{[10]}
\begin{equation}
A^b_j \; = \; - \, \frac{1}{2} \, F^b_{ji} \, x^i \quad ;
  \qquad A^b_0 \; = \; -\, F^b_{oi} \, x^i \quad .
\label{eq.12}
\end{equation}
The field $A_\mu$ is treated as an external field and therefore
satisfies the equation of motion $D_\mu F^A_{\mu\nu} = 0$, where
$F^A_{\mu\nu}$ is the background field strength which is assumed to be
practically constant over the extent of the heavy meson.

We next expand the nonrelativistic Lagrangian only to second order in
the quantum fields and subsequently integrate them out in favour of an
effective (coulombic) interaction. These calculations have been carried
out in more detail in ref.~{[9,11]} and will not be presented here. It is
important to mention that during the calculations retardation effects
have been neglected (this instantaneous approximation should be correct
up to order $1/m$) and that matrix elements of $A_\mu$ have been
parametrized, producing terms proportional to $\phi^2$. The resulting
effective Hamiltonian was presented in ref.~{[9,11]} and is still
complicated. We have numerically diagonalized it in the basis
(3 - 6) and found the solution of the resulting set of
coupled differential equations for the wave functions. We have then
checked that for the description of the low lying states of the
spectrum it is enough to keep the terms of order $(1/m)^0$ plus the
kinetic energy terms. The effective Hamiltonian can be finally written
as
\begin{eqnarray}
H & = & \int d^3 x \left\{ u^\dagger (\vec x \,) \, m \,
    u(\vec x \,) + v(\vec x \,) \, m \, v^\dagger(\vec x \,) \, - \,
      \frac{1}{4} \; F_{\mu\nu} \, F^{\mu\nu} \right.\nonumber \\[0.3cm]
   && - \, u^\dagger(\vec x \, ) \, T^A \, g \, E^A_i \, x_i \, u
    (\vec x \, ) - v(\vec x \, ) \, \overline T^A \, g \, E^A_i \,
     x_i \, v^\dagger(\vec x \, )  \nonumber \\[0.3cm]
   && - \, u^\dagger(\vec x \,) \; \frac{\vec\nabla^2_x}{2m}
       u(\vec x \, ) - v(\vec x \,) \; \frac{\vec\nabla^2_x}{2m}
                   \; v^\dagger(\vec x \,)  \nonumber \\[0.3cm]
&& +\, \alpha_s \int d^3 y \, u^\dagger(\vec x \,) \, T^A \,
    u(\vec x \, ) \, \frac{1}{r} \; v(\vec y \,) \,
     \overline T^A \, v^\dagger(\vec y \,)  \nonumber  \\[0.3cm]
\label{eq.13}
\end{eqnarray}
where the second line corresponds to the ``Stark effect'' discussed by
Leutwyler~{[14]} and Voloshin~{[13]}.
Here $u(\vec x\,)$ and $v(\vec x \, )$ denote the annihilation
operators for a quark and antiquark of mass $m$ respectively whose spin
and color indices have been suppressed, $\vec r = \vec x - \vec y$ and
$T^A$, $\overline T^A$ are the Hermitian generators of the $SU(3)$ color
Lie-algebra in the $3$ and $\overline 3$ representations respectively.

\eject

Diagonalysing our effective Hamiltonian in the basis~(3--6) only the
states $S$ and $E_1$ couple and we obtain the following set of coupled
differential equations for the wave functions.
$$
\left[2 m - E + C-\frac{1}{m}\;\frac{\partial^2}{\partial r^2}\,
    - \, \frac{4}{3} \; \frac{\alpha_s}{r}\right] S(r)\, =\,
       -\,\frac{\pi\phi\, r}{3\sqrt{2}}\; E_1(r)  \eqno (15 a)
$$
$$
\left[2m - E -\frac{1}{m}\left(\frac{\partial^2}{\partial r^2}\,-\,
      \frac{2}{r^2}\right) + \frac{\alpha_s}{6}\right]
         E_1 (r) \,=\, \frac{\pi \phi\, r}{3\sqrt{2}}\, S(r) \eqno (15 b)
$$
\noindent where $E$ is the mass eigenvalue of the quarkonium
and $\,m\,$ is the mass of the constituent quarks. The functions
$\,S(r)\,$ and $\,E_1(r)\,$ are related to the wave function
components in the expansion~(7) via
\[
\psi_S (r) = \frac{1}{r}\,S(r) \quad; \qquad
       \psi_{E_1} (r) = \frac{E_1}{r^2}\, (r) \quad .
\]

In order to include the scale
dependence (or distance dependence) of $ \alpha_s $ we use in eq. (15)
$$
\alpha_s=\alpha_s(r)=4\pi\frac{1}{b_0 f(r)}[1 + \frac{2\gamma_E+53/75}{f(r)}
- \frac{462}{625} \frac{ln(f(r)}{f(r)}]
$$
where
$$
f(r)=ln[\frac{1}{(\Lambda r)^2}+ b]
$$
$$
b_0=25/3
$$
This is the result of the two loop calculation given in ref.~{[12]}. In the
above expression $\gamma_E=0.5772$, $\Lambda=200$ MeV and $b =20$.
Among the matrix elements we find some involving the energy of the
background fields. In particular we find
$$
\langle S\,|\int\,d^3 x \left( - \frac{1}{4}\,F_{\mu\nu}\,
    F^{\mu\nu}\right) | S\rangle \,=\, \langle\Omega
     | F^2 | \Omega\rangle  \,=\,C\,=\,C_0 \phi^2  \eqno (16 a)
$$
$$
\langle E_1 | \int d^3 x \left( - \frac{1}{4}\, F_{\mu\nu}\,
     F^{\mu\nu}\right) | E_1\rangle \,\sim\,
        \langle \Omega|E F^2 E | \Omega\rangle = 0  \eqno (16 b)
$$

\noindent where $\phi^2$ is the
gluon condensate defined in (1) , $C_0$ is a positive constant and C is the
energy appearing in eq. (15a).
The last line follows from the assumption that higher order vacuum
expectation values of the background fields are zero.

As it can be seen there is a splitting between ``singlet'' and
``octet'' states given by C. Since $\phi^2$ is negative C will be also
negative and the singlet states have a negative energy with respect to the
octet states. The constant factor $C_0 $ was fixed by reproducing the
observed energy levels of the groundstate and first excited
charmonium and bottomium states at zero temperature. In the $(1/m)^0$
approximation pseudoscalar and vector mesons are degenerate and our
calculations are valid for the  $J/\psi$ , $\psi'$ ,$\Upsilon$ and
$\Upsilon'$ states. For an energy splitting between $\psi'$ and $J/\psi$
of $ 600$~MeV we obtain $ C=-737$~MeV.

In order to investigate the temperature dependence of our results we
consider
the temperature dependence of the gluon condensates. In view of the existing
estimates of this dependence we parametrize it in the folowing way :
$$
\frac{\phi^2}{\phi_0^2} = 1 - (\frac{T}{T_0})^4  \eqno (17)
$$
where $\phi_0^2$ is the value of the condensate at $T=0$ ( $\phi_0^2 = 360
MeV^4 $)
 and $T_0$ is some
critical temperature at which $\phi^2=0$. $T_0$ might be much larger than
the
deconfining transition temperature. With $T_0=250$~MeV eq.~(17) interpolates
the results compiled in ref.~{[6]}. Inserting eq.~(17) into (15) and solving
it for several values of T between 0 and $T_0$ we find the wave functions
and
masses of the fundamental and first excited states at different
temperatures.
 The results for the masses are shown in figure 1. The quark masses were
taken
to be $m_c = 1640$~MeV and $m_b = 4800$~MeV. The first interesting aspect in
figure 1 is that the masses are increasing with temperature. We understand
this
behaviour in the following way : any physical state considered is a mixture
of
singlet and octet components but for the low-lying states , such as $J/\psi$
, $\psi'$, $\Upsilon$ and $\Upsilon'$ the singlet component is largely
dominant.
In the first of equations 15 (15a) , the constant C has the effect of
shifting
the
energy of the state to smaller values (since it is negative). The
temperature
dependence of the condensate implies that
$$
C = C(T) = C_0 \phi^2(T)
$$
As $\phi^2(T)$ decreases with temperature , so does C and the energy levels
of
the states are shifted to larger values.
Roughly speaking we `` boil away ''
the physical vacuum and raise the energy of the states. A similar effect
occurs in the context of thermodynamics of quarks and hadrons during the
deconfining transition, where a suppression of the physical vacuum brings an
additional term (in the simple bag model language this is the bag constant)
to the quark energy density.

Our conclusion about the behaviour of heavy quarkonium masses with
temperature
is in contradiction with those of ref.~{[1]} and {[2]} but in agreement with
the lattice simulations performed in ref.~{[4]}.

The second interesting aspect in fig. 1 is the sudden disappearence of the
$\psi'$ line much below the critical temperature. Excited states (specially
light ones) are less tightly bounded and therefore sensitive to subtle
changes
in the confining potential. Moreover they contain larger octet components ,
which vary rapidly with the condensate.
 In our approach, as extensively discussed in
ref.~{[11]} , the gluon condensate generates the mid range part of the
potential,
which other authors parametrize as being linear. Our calculations indicate
that within the approximation 16b , i. e. , neglecting higher order
condensates,
a small reduction in $\phi^2$ due to the temperature is enough to make
the existence of $\psi'$ impossible. This might be true but might be
just a consequence of the approximation, which breaks down for very
large states.

There is some uncertainty in the quark masses $m_c$ and $m_b$. One might
think
that a different value of $m_c$ would change our results not only
quantitatively
 but also qualitatively. We therefore present in figure 2 plots of the
charmonium mass (scaled by its mass at zero temperature) as a function of
temperature for three different charm quark masses $m_1 = 1200$,
$m_2 = 1400$ and
$m_3 = 1640$ MeV. For each value of $m_c$ the octet-singlet energy splitting
constant C has to be chosen so as to properly reproduce the 1S-2S charmonium
mass splitting at T = 0. The chosen values were $C_1 = -995$ , $C_2 = -885$
 and
$C_3 = -737$ MeV. We concentrate on the charmonium because it is larger than
the bottomium and therefore more sensitive to variations of the parameters.
In figure 2a and 2b we show the behaviour with temperature
of the charmonium fundamental and
first excited state respectively. As it can be seen the curves corresponding
to the three quark masses are similar and exhibit the same increasing trend.
In the same way as charm states have a stronger dependence with temperature
than bottom states, we observe that lighter charm masses lead to a stronger
dependence of the bound state with temperature.
We have checked that changes in the values of the parameters $\Lambda$ ,
b and $\phi^2$ lead to curves with the same aspect of those presented in
figures 1 and 2. This is also true for the bottomium.
{}From what was said above we can see that our results are stable under
variations of the parameters.

\vskip10mm

\noindent
{\bf ACKNOWLEDGEMENTS}
\vskip 7mm

This work was partially supported by FAPESP -- Brazil and
 CNPq -- Brazil.
We are indebted to H. Leutwyler,to H. Satz and specially to M. Schaden
for fruitful discussions.
F. S. N. is grateful to H. Satz for the hospitality extended to him
in Germany , during the Workshop ``Thermodynamics of Quarks and
Hadrons''.

\eject
\noindent{\bf REFERENCES}
\begin{list}{}{\setlength{\leftmargin}{7mm}\labelwidth2.5cm
\itemsep0pt\parsep0pt}

\item[{[1]}] F.O. Gottfried and S.P. Klevansky, {\sl Phys. Lett.} {\bf
B286} (1992) 221.

\item[{[2]}] T. Hashimoto et al., {\sl Phys. Rev. Lett.} {\bf 57}
(1988) 2123; {\sl Z. Phys.} {\bf C38} (1988) 251.

\item[{[3]}] R.J. Furnstahl, T. Hatsuda and S. H. Lee, {\sl Phys.
Rev.} {\bf D42} (1990) 1744.

\item[{[4]}] T. Hashimoto et al., {\sl Nucl. Phys.} {\bf B400}
(1993) 267; {\sl Nucl. Phys.} {\bf B406} (1993) 325.

\item[{[5]}] S. H. Lee, {\sl Phys. Rev.} {\bf D40} (1989) 2484 ;
M. Campostrini and A. Di Giacomo, {\sl Phys. Lett.} {\bf B197}
 (1987) 403.

\item[{[6]}] V. Bernard and U.G. Mei$\beta$ner, {\sl Phys. Lett.} {\bf
B227} (1989) 465.

\item[{[7]}] J. Sollfrank et al., {\sl Nucl. Phys.} {\bf A566} (1994)
563c.

\item[{[8]}] S. Glazek and M. Schaden, {\sl Phys. Lett.} {\bf B198}
(1987) 42.

\item[{[9]}] C.A.A. Nunes, PhD. Thesis, T.V. Munich (1992).

\item[{[10]}] G. Curci et al., {\sl Z. Phys.} {\bf
C18} (1983) 135.

\item[{[11]}] C.A.A. Nunes, F.S. Navarra, P. Ring and M. Schaden, USP
Report IFUSP/P-1120, 1994.

\item[{[12]}] K. Igi and S. Ono, {\sl Phys.
Rev.} {\bf D33} (1986) 3349.

\item[{[13]}] M.B. Voloshin, {\sl Nucl. Phys.} {\bf B154} (1979) 365.

\item[{[14]}] H. Leutwyler, {\sl Phys. Lett.} {\bf 98B} (1981) 447.

\end{list}

\begin{figure}
   \begin{center}
      \leavevmode
      \epsffile{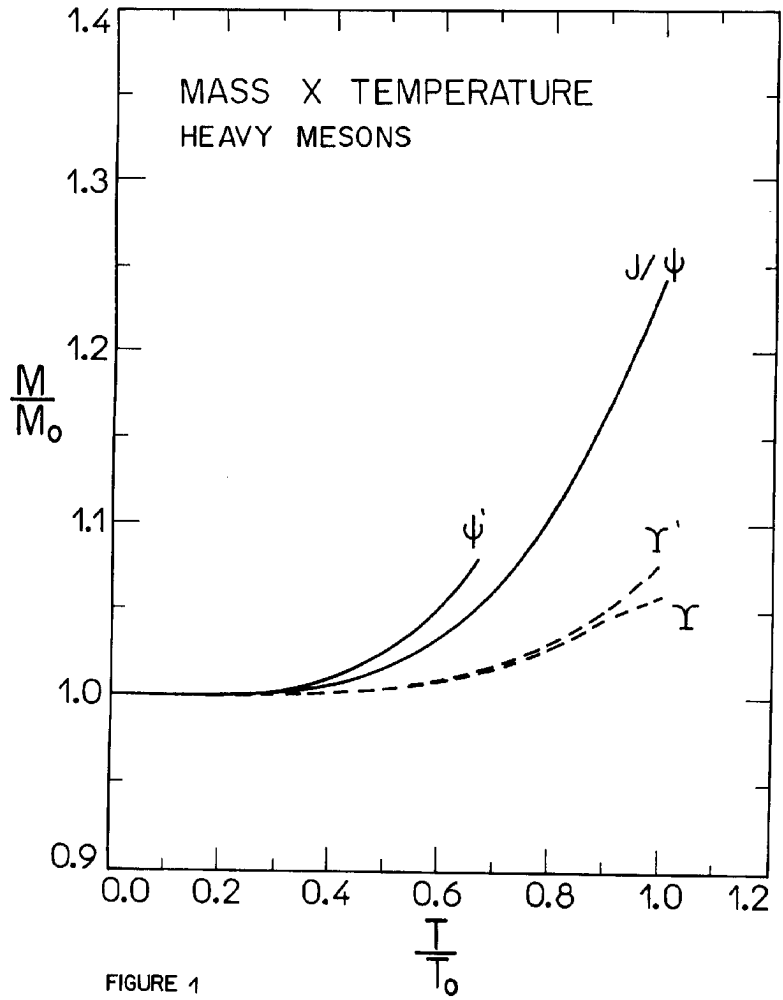}
   \end{center}
   \caption{ Mass (in vacuum mass units) plotted against
temperature (in critical temperature units). $m_c = 1640$~MeV}
\end{figure}

\begin{figure}
   \begin{center}
      \leavevmode
      \epsffile{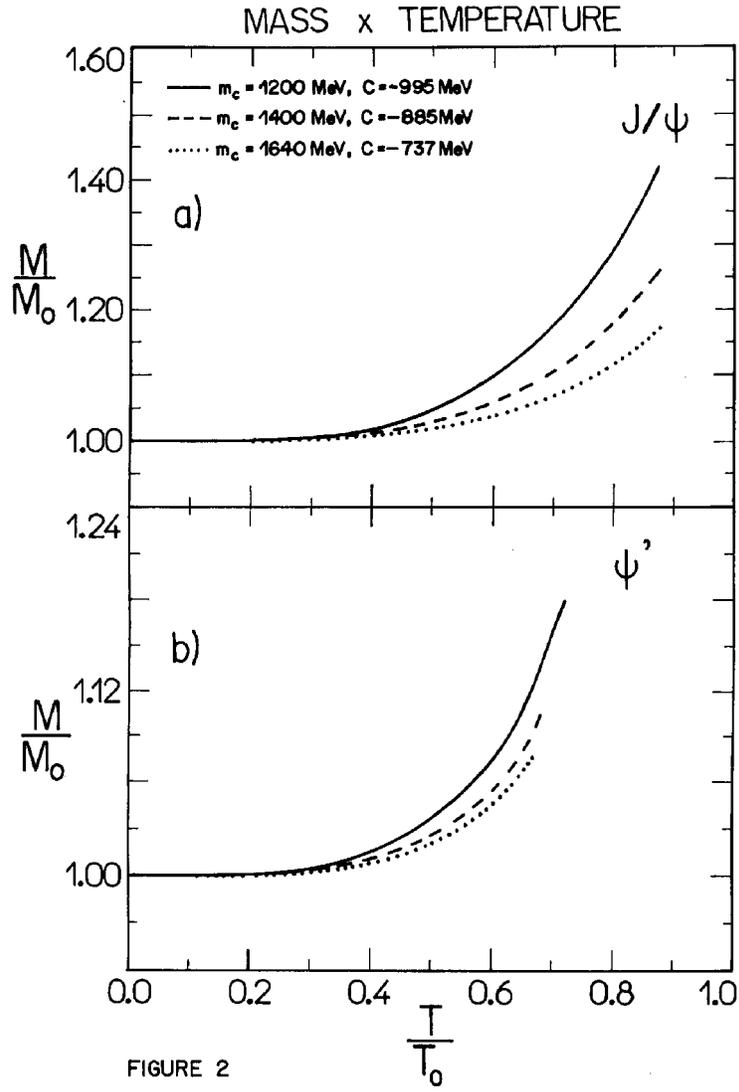}
   \end{center}
   \caption{ a) Mass (in vacuum mass units) of the charmonium
groundstate as a function of temperature (in critical temperature units)
for charm quark masses equal to 1200, 1400 and 1640 MeV. b) The same as a)
for the charmonium first excited  state.}
\end{figure}

\end{document}